\centerline{ \bf On the Four Dimensional Conformal Anomaly, Fractal Spacetime} 
\centerline{\bf and the Fine Structure Constant}
\bigskip
\centerline{ Carlos Castro}
\smallskip
\centerline { Center for Theoretical Studies of Physical Systems}
\centerline{ Clark Atlanta University, Atlanta, Georgia, 30314}
\bigskip
\centerline{ October 2000 }

\bigskip

\centerline{\bf Abstract}
Antoniadis, Mazur and Mottola ( AMM) two years ago 
computed the intrinsic Hausdorff dimension of spacetime at the 
infrared fixed point of the quantum conformal factor in $4D$ Gravity. 
The fractal dimension was determined by the coefficient of the Gauss-Bonnet 
topological term associated with the conformal gravitational anomaly and was found 
to be  greater than $4$. It is explicitly shown how one can relate the value of the 
Hausdorff dimension computed by AMM to the universal dimensional 
fluctuation of spacetime $\epsilon $ given by $\phi^3/2$, where $\phi$ is the Golden 
Mean $0.618..$. Based on the infrared scaling limit of the theory and using 
recent Renormalization Group arguments by El Naschie,  
we conjecture that the unknown coefficient $Q^2$, associated with the four 
dimensional gravitational conformal anomaly,  could be precisely equal to the 
inverse fine structure constant values ranging between $137.036 $ and $137.081$. 
Our results generate decimal digits up to any $arbitrary$ number.

\bigskip

\centerline {\bf The Conformal Anomaly and Fractal Spacetime at Large Scales } 

\bigskip 

Antoniadis, Mazur and Mottola [1] more than two years ago 
computed the intrinsic Hausdorff dimension of spacetime at the 
infrared fixed point of the quantum conformal factor in $4D$ Gravity. 
The fractal dimension was determined by the coefficient of the Gauss-Bonnet $topological$ 
term associated with the conformal anomaly : $trace~ anomaly$ and was found to be  greater than $4$. 
They also discussed a plausible physical mechanism for the 
$screening$ of the cosmological constant at very large distances 
in full agreeemnt with Nottale's work [7] and other results obtained by 
the present author and collaborators [6]. 

The Hausdorff dimension for spacetime is related to the geodesic distance 
$l(x, x')$ between points $x, x'$  and the volume $V_l$ enclosed by the 
spherical surface radius equal to $l$ . The scaling  relation between the two is 
$V_l \sim l^{d_H}$. For large $l$, this scaling relation 
defines the intrinsic dimension $d_H$ of the space. 

In $2D$ quantum gravity [2,3] the most appropriate way to calculate the $d_H$ is by the heat kernel methods 
associated with the Laplacian operator $D_\mu D^\mu$ and the proper time $s$ : 
$$ K_2 (x, x'; s, g ) = < x| e^{ - s D_\mu D^\mu } |x'> . \eqno (1) $$
The heat kernel $K_2$ has a short distance expansion whose anomalous scaling behaviour can be  
calculated based on the standard techniques pioneered by B. de Witt in the sixties. The average 
geodesic length-squared that a scalar particle is able to diffuse after a proper time $s$ is given by :

$$l^2_s \equiv {1 \over V} ~<~\int d^2x \sqrt g ~\int d^2 x' \sqrt {g'} ~ l^2 K_2 (x, x'; s, g ) ~>_V . 
\eqno ( 2 ) $$
where the average is taken with respect to the fixed volume Liouville field theory partition function. 
By expanding the heat Kernel $K_2( x, x'; s; g) $ in a power series $s$ one  can see that 

$$l^2_s \sim s ~~~ as ~~~ s \rightarrow 0. \eqno (3a) $$
which is the standard result undergoing Brownian motion.

The relevant scaling behaviour ( with area/volume ) 
is the expectation value of the following quantity appearing in the 
heat kernel expansion,  under the conformal scalings of the metric 
$g_{ab} = exp ( 2 \sigma)  {\bar g} _{ab}$ :

 $$s~<~\int d^2x \sqrt g ~ D_\mu D^\mu ~f_\epsilon (x, x_o)  ~>_V~   =  
s~<~\int d^2x \sqrt {{\bar g} } ~ {\bar D} _\mu {\bar D}^\mu ~{\bar f} _\epsilon (x, x_o)  ~>_V ~  
\sim s V^{{\alpha_{ -1} \over \alpha_1} } 
\eqno ( 3b) $$
where $f_\epsilon (x, x')$ is any smooth function with support only for distances of $l(x,x') \sim 
|x - x'|< \epsilon$. As $\epsilon$  goes to zero it approaches a delta function : 
$(1/\sqrt g) \delta^2 ( x-x')$

The finite area/volume scaling behaviour of the last proportionality factor follows by a constant 
shift in the Liouville field $ \sigma \rightarrow \sigma + \sigma_o $. The $ \alpha_n$ are the 
anomalous scaling dimensions associated with the fields  of the Liouville field theory 
and are given in terms of the weights  $n$ by the formulae :

$$\alpha_n = n + { \alpha^2_n \over Q^2 } = { 2n \over 1 + \sqrt { 1 - { 4n \over Q^2 } } } . \eqno (4a) $$
where the charge $ Q^2$ is determined in terms of the matter central charge ( anomaly coefficient ) 
$c_m$  by :

$$Q^2 = { 25 - c_m \over 6}. ~~~D = 2 .\eqno (4b) $$

We refer to the references [1,2,3 ] for further details. The main result is that the scaling behaviour 
of $s$ under a global area/volume scaling 
$$ s \rightarrow \lambda^{ - { \alpha_{-1} \over \alpha_1} } s  . \eqno(5) $$ 
will determine the Hausdorff Dimension $d_H$ from the relation :

$$ l^2_s \sim s \sim V_l^{ - { \alpha_{-1} \over \alpha_1} } \sim   
l^{ - d_H { \alpha_{-1} \over \alpha_1} }         . \eqno (6) $$
giving finally : 

$$d_H = -2 {  \alpha_1 \over \alpha_{-1} } = 2~ { \sqrt { 25 - c_m } + \sqrt { 49 - c_m } \over
  \sqrt { 25 - c_m } + \sqrt { 1 - c_m } }~  \ge 2.~~~ D = 2 \eqno (7) $$

Notice that the classical limit is obtained in eqs-(4a,4b,7) by seeting 
$ c_m = - \infty $ which implies that 
$Q^2 = \infty$ so that in eq-(7) the Hausdorff $ d_H = 2$ as it ought to. 
The authors [1] repeated this analyis associated with the conformaly anomaly in $ D = 4 $  where 
the charge $Q^2$ is now the coefficient of the Gauss-Bonnet curvature squared terms  
present in the four dimensional conformal anomaly :

$$ Q^2 = { 1\over 180} ( N_S + { 11\over 2} N_{WF} + 62N_V - 28 ) + Q^2_{grav} . \eqno (8) $$
the $unknown$ charge $Q^2$ is given in terms of the number of free scalars $N_S$ , Weyl fermions  $N_{WF} $ and vector
fields $N_{VF}$. While $ - 28$ and the $unknown$ value of the gravitational charge 
$Q^2_{grav}$ are the contributions of the 
spin-$0$ conformal factor  and spin-$2$ graviton fields of the metric itself. 

The scaling behaviour of the proper time under a global scaling of the volume $ V \rightarrow \lambda V$ in $ D = 4$ 
is : 

$$ s \rightarrow \lambda^{ - { \beta_8\over \beta_o   } } s . \eqno (9) $$
where $\beta_o, \beta_8$ are the conformal scaling exponents corresponding to the volume operator and the 
higher quartic derivative operator ( square of the Laplacian plus curvature and other derivative terms). 
The Hausdorff dimension is given, in the long scale limit, by the relation : 

$$l^4_s \sim s \sim V_l^{ - { \beta_8\over \beta_o   } } \sim 
l_s^{ -d_H  { \beta_8\over \beta_o   } } \eqno (10)$$
then the Haussdorff  dimension of fractal spacetime is explicitly given by :

$$d_H = - 4 {\beta_0 \over \beta_8} = 
4 ~  { 1 + \sqrt { 1 + {8\over Q^2 } } \over 1 +  \sqrt { 1 - {8\over Q^2 }}} ~   \ge 4. \eqno (11) $$
This final expression for the spacetime fractal dimension is {\bf all}  we need to show that the $unknown$ 
charge $ Q^2 $ in eq-(8) can be equated with the inverse of the fine structure constant $ 137.036....$ 
( given by the Particle data booklet ). 

The authors [1] emphasized that the value of $ Q^2$ was uncertain , principally because of  
the unknown infrared contributions of gravitons to the value of $Q^2_{grav} $ which appears in the r.h.s of 
eq-(8).  

To calculate what the value of $Q^2$ may be , which in turn will yield the value of 
$Q^2_{grav}$ present in the r.h.s of (8), we conjecture that this value can be related to the  
inverse of the fine structure constant $ 137.036..$ based on the recent papers by the author, 
Granik  and El Naschie [4,6] . The main results of [4] and [6] is that there is a 
$universal$ dimensional fluctuation in Nature given in terms of the Golden Mean by : 

$$ \Delta D_{ fluctuation}   = \epsilon = {\phi^3 \over 2} = \phi - { 1\over 2 } = 
(0.618 -{1\over 2 })  =
0.118..... ~ where~  \phi+1 = {1\over \phi} \Rightarrow \phi = { \sqrt {5}  - 1 \over 2 }= 
0.618....\eqno (12)  $$ 
and that the inverse of the fine structure constant [4] , ranging between $137.036...$ and El 
Naschie's results $137.081..$, 
can be thought of as an $internal$ dimension or a $central$ charge, using the 
language of Irrational Conformal field Theory,
as Eddington envision long ago. This line of reasoning is nothing but following the path 
chartered by Einstein himself on the geometrization of $all$ physics, with the new ingredient that 
we believe that Nature {\bf is}  fractal at its core.

A simple numerical calculation shows that by simply  setting $ Q^2 = 137.036.... $ 
inside  the basic equation (11) yields automatically nothing more, nothing else but $ 4 +\epsilon $ 
for the Hausdorff dimension of fractal spacetime , where 
$\epsilon $ is the universal dimensional fluctuation given by $\phi^3/2$ :

$$4  ~ { 1 + \sqrt { 1 + {8\over 137.036 } }  \over 1 +  \sqrt { 1 - {8\over 137.036 } }  }  = 4.11856 \sim 
4 + { \phi^3 \over 2} = 4.118.....~~~~ !!!!!!. \eqno (13)$$

Is this a {\bf numerical concidence} or {\bf design } ????. Conversely, we can propose  an {\bf exact} value 
for the inverse of the  fine structure constant simply by setting the right hand side of ( 13) to be precisely 

$$d_H = 4+ \epsilon = 4 + \phi^3/2 = 4 + { \sqrt {5}  - 2 \over 2 } \equiv    
4  ~ { 1 + \sqrt { 1 + {8\over Q^2 } }  \over 1 +  \sqrt { 1 - {8\over Q^2  } }  }. \eqno ( 14 )    $$
Eq-( 14 ) will determine the $precise$ value of $Q^2 = 137......$ 
up to an $arbitrary$ number of digits. This is very encouraging because the only thing we have to 
do is to compare the value we get for the inverse fine structure constant, $Q^2$, 
 and compare with the results of {\bf QED} !!!
This could very well be the {\bf sought-after} $ numerical $ proof of the correctness of the theory. 
Evenfurther, notice that in this infrared limit, we are working entirely with ordinary $4D$ gravity 
( long distance limit of string theory ) and there is no need to do any compactifications 
from higher to lower dimensions.

The graph of $ d_H = d_H ( Q^2)$ is a decreasing function of $Q^2$ 
( the internal dimension in Eddington's language ) . When $ Q^2 \rightarrow \infty$ the $ d_H = 4$ and one 
recovers the classical four dimensional limit. When $ Q^2 < 8$ the dimension becomes $ complex $ 
which is not so a farfetched thing to imagine. One can envision an analytical continuation of the dimension 
consistent with the complexification of ${\cal E}^{ (  \infty) } $ spacetime 
and the celebrated Banach-Tarski 
theorem. One can imagine a sort of Quantum/Classical $duality$ as  follows : 

Earlier in eqs-(4a,4b,7) we discussed  that the classical limit is recovered when the 
matter central charge ( or the embedding  target spacetime dimension using the language of string theory ) 
is $ c_m = d = - \infty  ; ~ Q^2 = \infty $ so that the Hausdorff dimension of the $worldsheet $ is now 
$ d_H = 2$. 
This point $ d  = -\infty$ has for $antipodal ~ point$ the one of  $ d  = \infty$ 
which correspond, respectively,  to the topological dimensions associated with the totally 
$void$ set $ {\cal E}^{ ( - \infty )}$ 
and the total space $ {\cal E}^{ (  \infty )} $ . The former has for Hausdorff  dimension the value of $0$ 
and the latter has for Hausdorff dimension $\infty$. It is clear now why  the void space 
$ {\cal E}^{ ( - \infty )} $ is the dimension-dual space to $ {\cal E}^{ (  \infty )} $ 
since $ \infty = 1/0$. 
In this sense, we can argue that there a sort of  Quantum/Classical duality principle operating. 

The Banach-Tarski theorem, which roughly states that you can get two objects out of one , 
all of them identical, by abandoning the Axiom of Choice in Set Theory, 
fits very well with the $complexification$  of  $ {\cal E}^{ (\infty )} $ space by having 
another $dual$ space : $ {\cal E}_{dual}^{ (\infty )} $ or  $mirror$ image. 
One can then perform the analytical continuation ( or duality quantum/classical transformation ) 
from $ d = \infty $ to $d = - \infty$, by moving along a large circle ( infinite radius ) 
in the complex-dimensions  plane , and reach $ d = - \infty$ from  $ d = \infty$. 
And then start over again and again  moving in infinite cycles . 
The quantum $vacuum$ , at $ d = \infty$,  
 is in this picture both a $source$  and a $sink$. The creation of our Universe out of 
`` nothing `` could have occurred in this fashion. 
This resembles the picture of the Snake (  coined $ouraborus$ ) 
eating its tail, then re-emerging again out of the vacuum and eating its tail later on  
and beginning another cycle, again and again, ad infinitum.

After this slight detour on the nature of $complex$ dimensions, we conclude by saying that 
El Naschie [4] has presented very convincing arguments regarding the unfication of gravity with the 
electroweak and strong forces, based on Renormalization Group Arguments, 
supporting that there is a very deep and  explicit connection bewteen
Nottale's Scale Relativity [7] , Irrational Conformal Field theory, 
El Naschie's Cantorian-Fractal spacetime and the 
New Extended Scale Relativity formulated by the author [5], and   
that the inverse fine structure constant ( an internal dimension or charge ) 
plays a fundamental role in determining the 
scaling regimes of the electroweak and strong interation as well as  
the hierarchy of the $16, 6 $ internal dimensions,  present in the Heterotic string 
theory and its compactifications , from $ 26 \rightarrow 10 \rightarrow 4 $ . 
Argyris et al [8] have recently 
shown how a fractalization of spacetime may be an intrinsic property of all processes in Nature, 
from the microworld to cosmos, having well defined signatures in 
Cosmic Strings and in the phenomenon of Spontaneous Symmetry breaking .

\centerline { Acknowledegements }
\smallskip
The author thanks M. S. El Naschie for his hospitality 
and for the series of discussions which led to this work.
Also we wish to thank S.Ansoldi for his assitance and to E. Spallucci, E. Gozzi, M. Pavsic,  
A. Granik, T. Smith, C. Handy. A. Schoeller, A. Boedo, J.Mahecha, J. Giraldo, L. Baquero  
for their support. 

\bigskip

\centerline {\bf References }

1. Antoniadis, P. Mazur and E. Mottola : `` Fractal Geometry of  Quantum Spacetime at Large Scales `` 

hep-th/9808070 . 

 Antoniadis, P. Mazur and E. Mottola : Nuc. Phys. {\bf B 388  } (1992 ) 627

2- V. Knizhnik, A. Polyakov and A. Zamolodchikov : Mod. Phys. Lett {\bf A 3 } 
(1988 ) 819.

F. David,  Mod. Phys. Lett {\bf A 3 } (1988) 1651. 

F. David : Nuc. Phys. {\bf B 257} (1985) 45  

3- J. Distler, H. Kawai : Nuc. Phys. {\bf B 257} (1985) 509 

V. Kazakov, Phys. Lett { \bf B 150} ( 1985) 282 

L. Ambjorn, B. Durhuus, J. Frohlich and P. Orland :  Nuc. Phys. {\bf B 270 } (1986) 457 . 

4- M. S. El Naschie : `` Coupled oscillations and mode locking of Quantum Gravity fields,
Scale Relativity and ${\cal E}^{(\infty)}$ space ``.   
Chaos, Solitons and Fractals {\bf 12} (2001) 179-192. 

5-C. Castro : `' Hints of a New Relativity Principle from $p$-Brane Quantum Mechanics `` 
Chaos, Solitons and Fractals {\bf 11} (2000) 1721  

C. Castro : `` Noncommutative Geometry, Negative Probabilities and 
Cantorian Fractal Spacetime `` Chaos, Solitons and Fractals {\bf 12} (2001) 101-104 

6-C. Castro, A. Granik : `` Scale Relativity in  ${\cal E}^{(\infty)}$ space and the Average Dimension of the World ' To appear in 
Chaos, Solitons and Fractals. hep-th/0004152

7- L. Nottale : `` Fractal Spacetime and Microphysics : Towards a theory of Scale 
Relativity. World Scientific , 1993. 

L. Nottale : `` La Relativite dans tous ses Etats `` Hachette Literature, Paris 1998. 

8- J. Argyris, C. Ciubotarium H. Matuttis :  `` Fractal space, cosmic strings and spontaneous 
symmetry breaking ``  
Chaos, Solitons and Fractals {\bf 12} (2001) 1-48

\bye